\documentclass{aa}
\usepackage{psfig}
\usepackage{graphicx}

\newcommand{\be}{\begin{displaymath}}
\newcommand{\ee}{\end{displaymath}}
\def\ros{{\sl ROSAT}}
\def\chan{{\sl Chandra}}
\def\pdot{\dot{P}}
\def\fdot{\dot{f}}
\def\fddot{\ddot{f}}
\def\edot{\dot{E}}
\def\bdot{\dot{B}}

\def\ns{RBS 1223}
\def\ins{isolated neutron star}

\def\Xray{\mbox{X-ray}}
\begin{document}

\title{Discovery of 5.16~s pulsations from the isolated neutron star RBS 1223}

   \author{V.\,Hambaryan\inst{1} \and
	   G.\,Hasinger\inst{1} \and
           A.D.\,Schwope\inst{1} \and 
	   N.S.\,Schulz\inst{2}}

   \offprints{V. Hambaryan, vhambaryan@aip.de}
   
   \institute{Astrophysikalisches Institut Potsdam,
              An der Sternwarte 16, D-14482 Potsdam, Germany
\and Massachusetts Institute of Technology, Center for 
Space Research, 70 Vassar Street, Cambridge, MA 02139, USA}

   \date{Received; accepted}
   
\abstract{
The isolated neutron star candidate \ns\
was observed with the Advanced CCD Imaging Spectrometer 
aboard the \chan\ \Xray\ observatory on 2000 June 24.
 A timing analysis of the data yielded a periodic modulation with 
a period 
$P=5.1571696^{+1.57\times 10^{-4}}_{-1.36\times 10^{-4}}$~s.
Using \ros\ HRI archived observations we  detected a period
$P=5.1561274\pm 4.4\times 10^{-4}$~s and determined period derivative
$\pdot=(0.7 - 2.0)\times 10^{-11}$
~s~${\rm s^{-1}}$.
The detection of this period and $\pdot$ indicates that \ns\ 
has a ``characteristic'' age of 6000 $-$ 12000 years and 
huge magnetic field at the surface 
($B_{dipole} \approx (1.7- 3.2) \times 10^{+14} {\rm G}$) 
typical for anomalous \Xray\ pulsars (AXPs).
}
\authorrunning{Hambaryan et al.}
\titlerunning{Discovery of 5.16~s pulsations from the RBS 1223}
   \maketitle

      \section{Introduction}\label{Intro}

Neutron stars (NS) supposedly represent a non$-$negligible fraction  of 
$\sim 1 \%$
of all stars in the Galaxy, with a total number as high as $10^8-10^9$ 
(e.g. Narayan \& Ostriker \cite{naost}).
Isolated NS represent the bulk of the population and soon after their
 birth may appear as radio pulsars. 
This pulsar phase could last some $10^7$~yr and these
NS are observable, if the beaming conditions are favorable. 
At formation NS are very hot 
($T \sim 10^{11} {\rm K}$), and they may give off thermal radiation while 
cooling.   

Depending on some model assumptions, a few hundred  
to several thousand accreting isolated NSs were expected to be 
detectable  in 
the ROSAT All-Sky Survey (Treves \& Colpi \cite{trevescolpi}, 
Blaes \& Madau \cite{blaesmadau}, Neuh\"auser \& Tr\"umper \cite{ralphT}). 
However, to date only 6--7 good candidates (Treves et al. \cite{trevesetal}, 
Motch \cite{motch2000}) for isolated NSs were discovered in ROSAT data 
(RX\,J1856.6--3754, Walter \cite{walter96}, 
 RX\,J0720.4--3125, Haberl et al. \cite{haberl97}, 
 RX\,J0806.4--4123, Haberl et al. \cite{haberl98}, 
 1RXS J130848.6+212708$\equiv$ RBS1223, Schwope et al. \cite{schwope99}, 
 RX\,J1605.3+3249, Motch \& Haberl \cite{motch99} and 
 RX\,J0420.0--5022, Haberl et al. \cite{haberl99}).
 They can be exclusively described by their similar \Xray\ properties, 
i.e.  they have 
soft \Xray\ spectra  which 
are well represented by pure blackbody emission with 
temperatures $kT $ between 50 and 120 eV, and extremely high \Xray\ to 
optical flux ratios, log($f_{x}/f_{opt}$) of $4-5.5$ 
(see tables in  Treves et al. \cite{trevesetal}, Motch \cite{motch2000}).

Most probably, these best candidates constitute the bright end of the 
log N$ - $log S distribution of \Xray\ detected isolated NSs 
(Neuh\"auser \& Tr\"umper \cite{ralphT}), suggesting
that the smaller number of  observed accreting old neutron stars 
than those theoretically expected
is mainly caused by their larger space velocities compared to 
the  previously assumed velocity distribution.  
This conclusion is supported  by the recent
discovery of a proper motion for RX\,J1856.6--3754 (Walter \cite{walter01}).
 Two of these objects (RX\,J0720.4--3125, RX\,J0420.0--5022) exhibit 
pulsations with remarkably long periods of 8.39~s and 22.7~s. 

The point source \ns\ (1RXS J130848.6+212708) was detected in the 
ROSAT All-Sky Survey 
(1RXS catalogue, Voges et al. \cite{vogesetal}) and studied as a candidate 
 \ins\ (Schwope et al. \cite{schwope99}) during the program of optical 
identifications of bright($> 0.2$ PSPC cts s$^{-1}$), 
high-galactic-latitude ($|b|>30^o$) \Xray\ sources (ROSAT Bright Survey -- RBS,
Schwope et al. \cite{schwope00}). 

A pointed \ros\ HRI observation (Schwope et al. \cite{schwope99}) determined
the position of the source to be $\alpha {\rm (J2000)}=13^h08^m48.17^s, 
 \delta {\rm (J2000)}=21^o27'07.5\arcsec (\pm 1.6 \arcsec)$. 
The corresponding Galactic coordinates are $l=339^o,b=+83^o$. 
Deep optical observations  with Keck-II failed to detect an 
optical counterpart down to a limiting magnitude of  $m_B \sim 26$, thereby 
placing a lower limit of $\sim 10000$ on the \Xray\ to optical flux ratio
(Schwope et al. \cite{schwope99}).

In this paper we report the detection of pulsations from the
isolated neutron star \ns\ based on recent observations  with 
the \chan\ \Xray\ observatory and \ros\  archival  observations.

\section{Observations and Data reduction}

The point source \ns\ was observed with the Advanced CCD Imaging 
Spectrometer (back illuminated S3 chip)  on board the \chan\ X--ray 
observatory on 2000 June 24, for a total of $\sim 10~{\rm ksec}$.
The source was placed at the farthest corner of 
the back-illuminated CCD S array,
close to the read out of node 3 in order to mitigate pile--up and charge 
transfer inefficiency effects.  
Only a small part (1/8) of the CCD was readout to additionally avoid pile--up 
and increase the time resolution to $\sim 0.5~{\rm s}$.

We have used level 2 ACIS event data, retaining only
'standard' grades 0,2,3,4 and 6  thus discarding events with very large
pulse heights as due to cosmic rays.

An inspection of an  X-ray image in chip coordinates (rather than sky 
coordinates in which the dithering of space-craft has been accounted for) 
showed no 'bad' pixels or columns in the source data.
Examination of the time dependence of the background did not reveal 
periods of high background during the observation.
 
To identify point sources in the field (Fig.~\ref{image}), 
we  used the routine CELLDETECT  provided by 
the Chandra Interactive Analysis of Observations (CIAO 2.0) 
software package. We determined the position of \ns\ 
to be $\alpha {\rm (J2000)}=13^h08^m48.26^s, 
\delta {\rm (J2000)}=21^o27'06.75\arcsec$, compatible with  the
position found in previous \ros\ HRI observations 
( see Sect.~\ref{Intro}). 

\subsection{Timing analysis}

\subsubsection{\chan\ \Xray\ observatory ACIS$-$S}

\begin{figure}
\psfig{figure=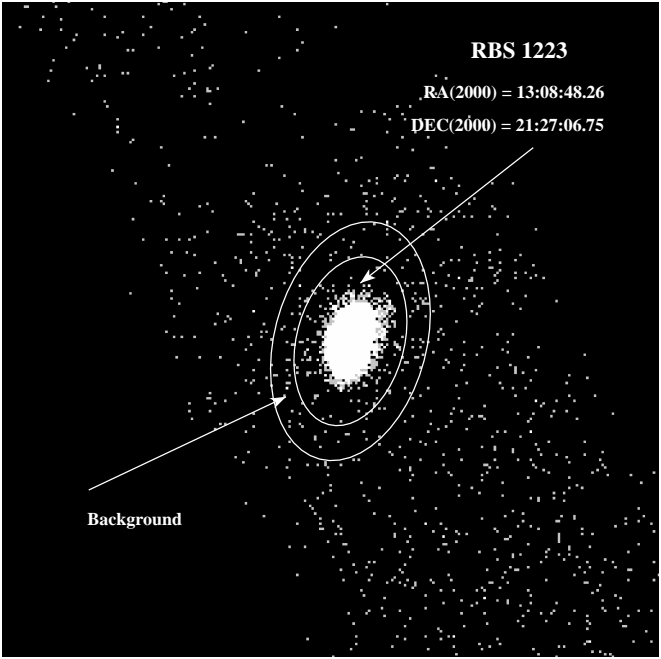,width=8.8cm,clip=}
\caption[ ]{Chandra ACIS-S image of RBS 1223. Inner ellipse corresponds
to the 5~$\sigma$ radius of the source (available from source detection 
procedure, CIAO routine CELLDETECT) and elliptical annuli to the background 
covering an equal area.}
\label{image}
\end{figure}

First of all, we performed a timing analysis of \chan\ \Xray\ 
observatory data  in order to search for periodic modulations. 
We extracted $\sim 7840$ ACIS--S source photon events 
(Fig.~\ref{image}), which we corrected for barycenter 
using the ``axBary'' programme, available at $ftp://asc.harvard.edu$. 

An application of various methods (e.g. Scargle \cite{scargle}, 
Buccheri et al. \cite{zsqr}) for periodicity search indicates
a highly significant detection of a periodic signal at 5.157~s. 

To evaluate the pulsation frequency more precisely,
and find its uncertainty, we employed the  Gregory-Loredo (GL) 
Bayesian method (Gregory \& Loredo \cite{gl92}, \cite{gl93}, \cite{gl96}).
This method is designed for the detection of a periodic signal 
of unknown shape and period, assuming  a Poisson sampling distribution. 
To compare  different models (how the data favor a periodic 
model of a given 
frequency $f$ with $m$ phase bins over the constant rate model) 
it calculates a frequency-dependent Bayesian odds ratios ($O_m(f)$)
(for details, see Gregory \& Loredo \cite{gl92}). 

\begin{figure*}
\psfig{figure=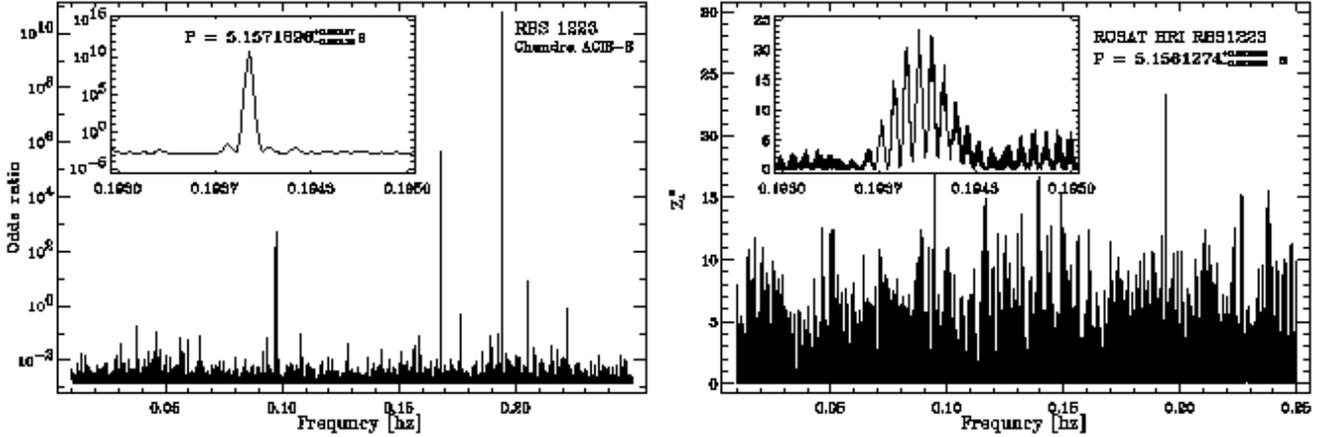,width=17.6cm}
\caption[ ]{Periodogrammes of RBS 1223: Chandra ACIS-S ({\it left}) 
and ROSAT HRI ({\it right}) observations}
\label{prdgrms}
\end{figure*}

We implemented the GL method, taking  the maximum 
number of phase bins equal to
15 and  surveyed the frequency range 0.01--0.25 Hz.
The odds ratio as a function of frequency is shown 
in Fig.~\ref{prdgrms} (left panel). As  a most probable 
value for the period we took the mode  
of the posterior probability distribution function (PDF).

The maximum value, $O_{\rm m}^{\rm max}=6.16\times 10^{10}$,
is  at $f_0=0.19390681$~Hz.
We computed the uncertainty of $f_0$ at 68\%
confidence level (``posterior bubble'') using the full Bayesian 
posterior PDF. It is equal to 
${\delta}f=^{+3.103}_{-7.913} \mu {\rm Hz}$.
Thus, we finally derive
the period of the detected pulsations  and the uncertainties to be 

\be
P_{Chandra}=5.1571696^{+1.57\times 10^{-4}}_{-1.36\times 10^{-4}} \;\; {\rm s}.
\ee

at the epoch of 51719.949676196 MJD.

The light curve extracted at $f=f_0$ (Fig.~\ref{lces} left panel)
reveals one broad pulse per period. The semi-amplitude modulation,
\be
A_{Chandra} \equiv \frac{CR_{max}-CR_{min}}{CR_{max}+CR_{min}},
\ee
where $CR$ is the count rate, is equal to $23\% \pm 5\%$.
Assuming that the detected signal is sinusoidal  
we estimate that the pulsed fraction  intrinsic to the source is
$20\%   \pm  2\%$ (Brazier \cite{braz}).

\subsubsection{\ros\ HRI}

The field of RBS 1223 has also been observed with the \ros\ HRI on 
June 13, 
1997 (ROR 703348) and on January 10, 1998 
(ROR 704082) for a net exposure time of 2218~s and 4938~s, 
correspondingly.

We also performed a timing analysis of the \ros\ HRI observations.
In both cases we used a $12^{\arcsec}$ radius circular region to extract the 
 source photons. 
This radius encompassed 85\% 
of the events in the \ros\ point spread function.  The extracted 
data sets consist of 266 and 521 counts, respectively. 
Photon arrival times were
solar barycenter corrected, as implemented in 
the Extended Scientific  Analysis Software System 
(Zimmerman et al. \cite{zimm}). 

We analyzed the $\sim 266$ HRI (ROR 703348)
photons from \ns\  for periodicities in the \Xray\ flux, but did not 
detect any significant signal using a Rayleigh $Z_1^2$ test 
(Buccheri et al. \cite{zsqr}). However, a similar analysis of the second
 data set (ROR 704082, consisting of 521 photons) 
revealed a periodic modulation at 0.19394401~Hz
(Fig.~\ref{prdgrms} right panel). 

As Fig.~\ref{prdgrms} shows, the significance of detection is not very high 
($Z_1^2 {\rm = } 23.2 $) owing to  the small number of registered photons. 
Moreover, the power spectrum is quite complicated;  it consists of 
a central peak 
accompanied by many side lobes owing to the data gaps.
 This makes the computation of a credible region  difficult, since 
the posterior PDF 
obtained by an application of the GL method
is not unimodal, unlike in the case of the \chan\ observations. 
We determined the uncertainty of this period measurement
of \ns\ through simulations, based on the 
measured pulsed fraction $\sim 0.3$ and count rate $\sim 0.1$s$^{-1}$. 
Each of our 10000 simulated data sets
(photon arrival times) was generated from a sinusoidal light curve and 
 obeyed Poissonian statistics. Each simulated data set had the 
same characteristics as
the observed data in terms of number of registered counts, spanned time, 
good time intervals, data gaps, and was analyzed exactly in the same way. 
The dispersion in the obtained period distribution was adopted as an 
estimate of the uncertainty in our measurement of the period of RBS 1223:
$\delta P = \pm 3.949\times 10^{-5}$~s.

 An error estimate can be derived alternatively from a more general 
approach
In general, the accuracy  of a frequency determination in a periodogram 
depends on the total observation time $T_{total}$, 
the number of measurements $N$ and the
signal-to-noise ratio (see e.g. Jaynes~\cite{jaynes}, 
Bretthorst~\cite{bretth}, Kovacs~\cite{kovacs}).
 
 According to these authors, the theoretical accuracy for determination
of the frequency of a single 
steady sinusoid ($ d_i = A \cos{(2\pi f t_i)} + e(t_i)$, 
single sine wave plus noise e(t$_i$) of standard deviation $\sigma) $ is

\be
  \delta f = \frac{1.1\sigma}{A\cdot T_{total} \cdot \sqrt{N}} {\rm Hz,}
\ee

\begin{figure*}
\hbox{\hspace{0cm}\psfig{figure=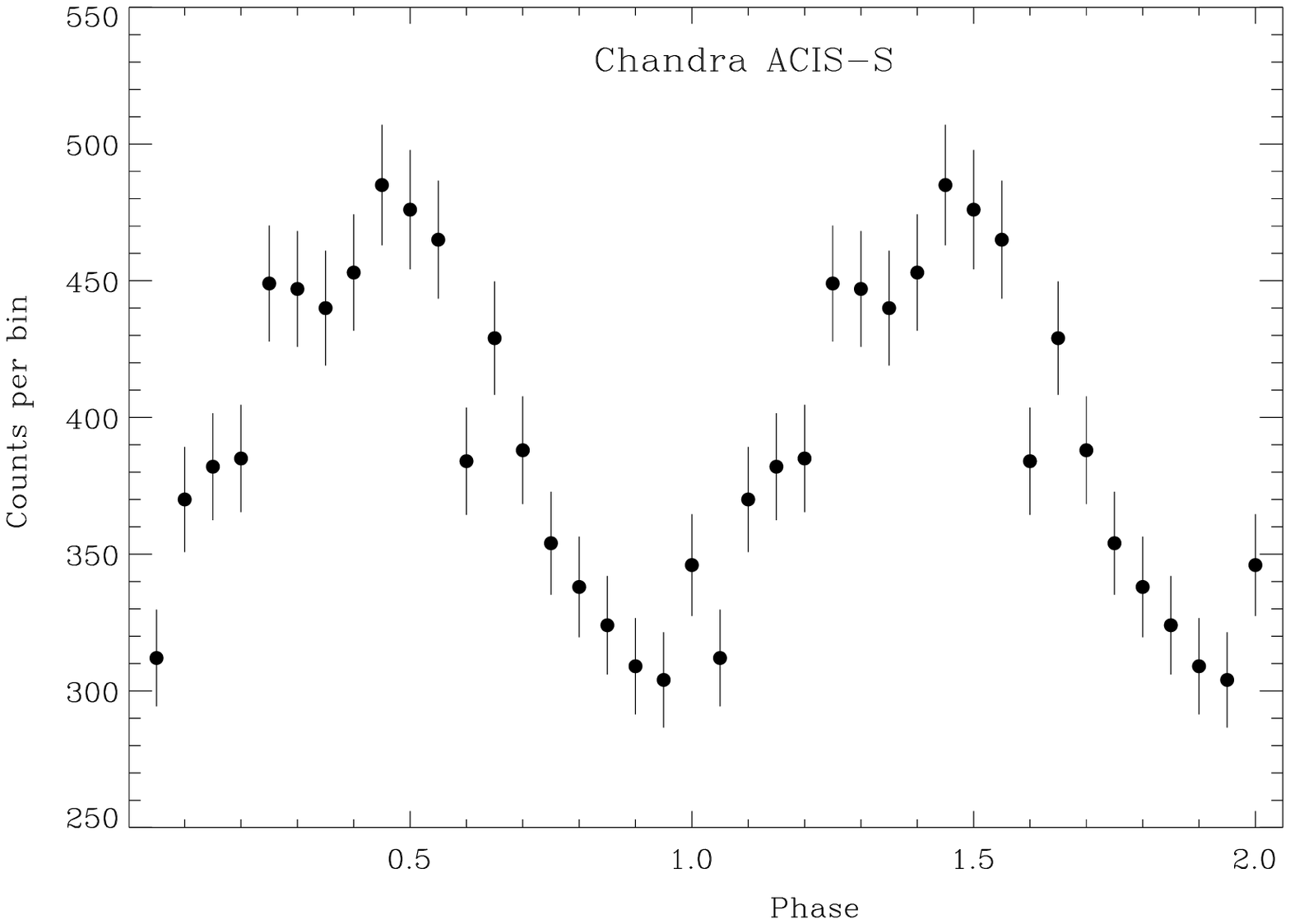,width=8.8cm}\hspace{0cm}
\psfig{figure=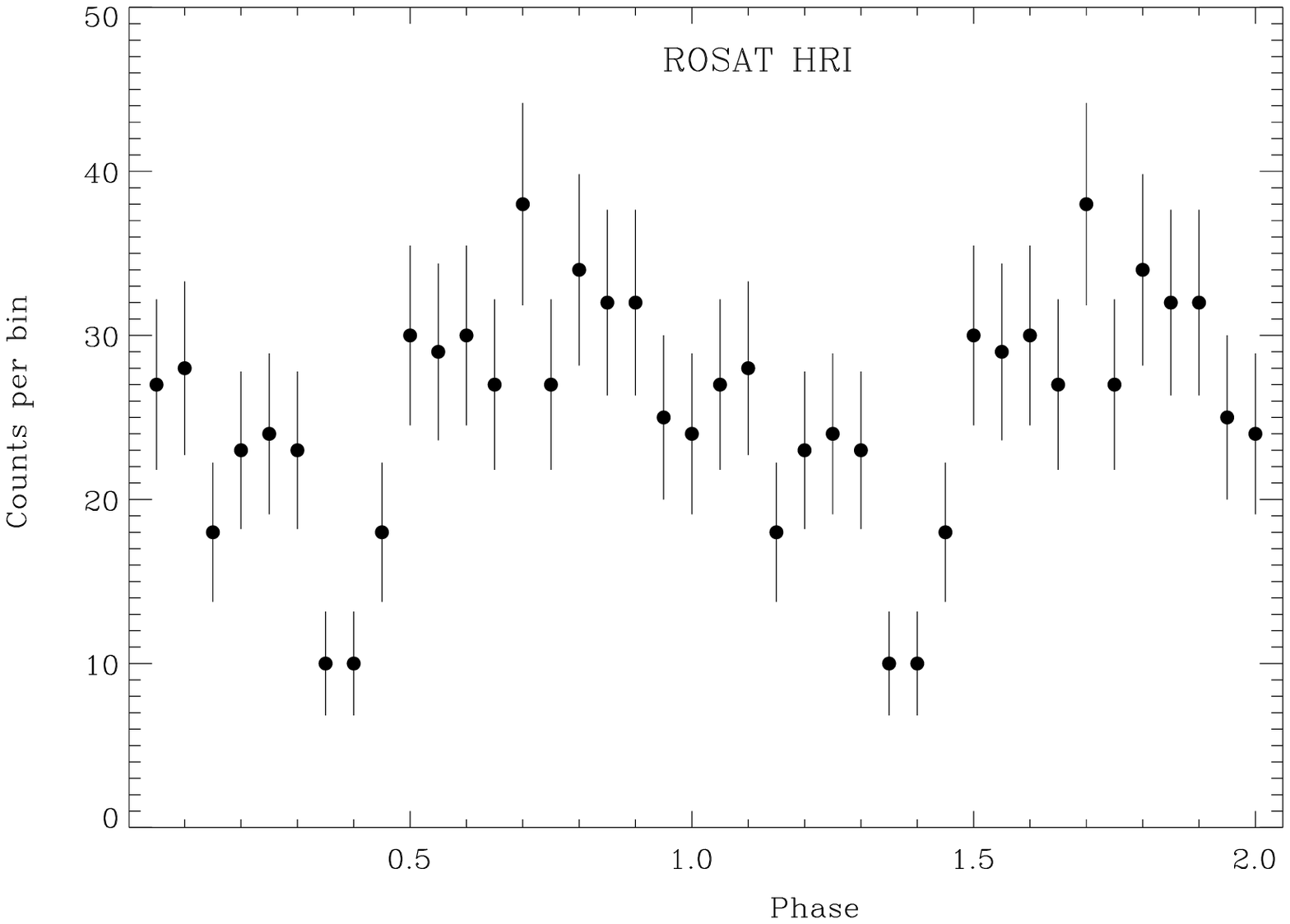,width=8.8cm}}
\caption[ ]{Phase folded light curves of RBS 1223: Chandra ACIS-S ({\it left}) 
and ROSAT HRI ({\it right}) observations}
\label{lces}
\end{figure*}

Hence, for the \ros\ observation of RBS 1223 (ROR 704082)  we get: 
$T_{total} \approx 70 $~ksec, N = 20 
(number of bins in the phase folded light curve)  and signal-to-noise 
ratio $A/ \sqrt{2} \sigma$~$\sim$ 3 (see Fig.~\ref{lces}, right panel). This 
leads to an optimistic estimate of accuracy of 
the period of:

\be
    \delta P_{optimistic} = \pm 4.4 \times 10^{-5}~s,
\ee

while a pessimistic  estimate is:

\be
\delta f = \frac{1}{2T_{total}} \;\; {\rm Hz~(Rayleigh~resolving~power) } 
\ee

\be
  \delta P_{pessimistic} = \pm 3.8 \times 10^{-4} {\rm ~s.}
\ee

Thus, we finally derive the period of the detected pulsations in the 
\ros\ HRI observations 

\be
P_{ROSAT}=5.1561274 \pm 3.8\times 10^{-4} \;\; {\rm s}
\ee

at the epoch of 50824.311057318  MJD.

This determination allows to calculate a period derivative
and its uncertainty

\be
\pdot= 1.35^{+0.69}_{-0.67} \times 10^{-11} \;\; {\rm s s^{-1},}
\ee

if we adopt  the pessimistic error estimate of the accuracy 
for period determination of RBS 1223 of \emph{ROSAT\/} 
HRI observations ($\delta P_{pessimistic})$.

\subsection{Spectral analysis}

Since the \chan\ \Xray\ observatory ACIS--S detectors have intrinsic energy 
resolution and the source spectrum has sufficient number counts 
(count rate $\sim$  $0.82\pm 0.02$~cts~s$^{-1}$), we were able to examine 
the spectral properties of RBS 1223. After extracting
(CIAO tool ``dmextract'') the spectrum of \ns\ and the nearby background,
we fitted the spectrum with an absorbed blackbody model using the 
XSPEC V11.0 package.

The ACIS--S instrumental response is not yet known in 
the energy range $< 0.45~{\rm keV}$ (see caveats page at 
$http://asc.harvard.edu/cda/caveats.html$).
 Hence, the absorption column density ($N_{\rm H}$)  cannot be well
constrained by the Chandra observation.
The \ros\ PSPC spectra based on All-Sky Survey observation extends down
to the 0.1~keV.
We therefore in our fitting analysis first of all used only \ros\ PSPC data
in order to determine $N_{\rm H}$ and then fixing that value 
to fit \chan\  ACIS--S data in the energy range 0.45--1.5 keV, 
since no photons above 1.5~keV were detected. 

\begin{figure*}
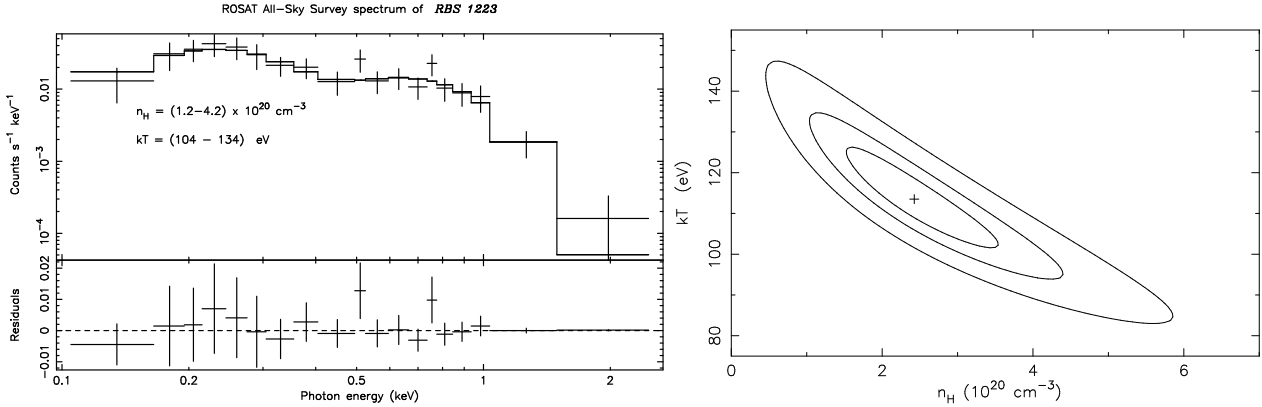

\hbox{\hspace{0cm}
\psfig{figure=fig4a.ps,width=8.8cm,angle=-90,clip=}\hspace{0cm}
\psfig{figure=fig4b.ps,width=7.8cm,angle=-90,clip=}}
\caption[ ]{A blackbody fit  
and  68\%, 90\% and 99\% 
confidence contours ($\Delta \chi^2$ = 3.5, 6.3, and 11.3, respectively) 
to the ROSAT PSPC spectrum of RBS 1223}
\label{specs}
\end{figure*}

The extracted spectrum and corresponding fit of \ros\ All-Sky Survey 
observation is shown in Fig.~\ref{specs}. 
The best fit parameters are a column density  of 
$N_{\rm H} = (2.43 \pm 1.06) \times 10^{20}$~cm$^{-2} $, 
and a temperature $kT_{BB} = (113.7 \pm 11.8)$~eV with formal 
1~$\sigma$ errors.

\begin{figure}
\psfig{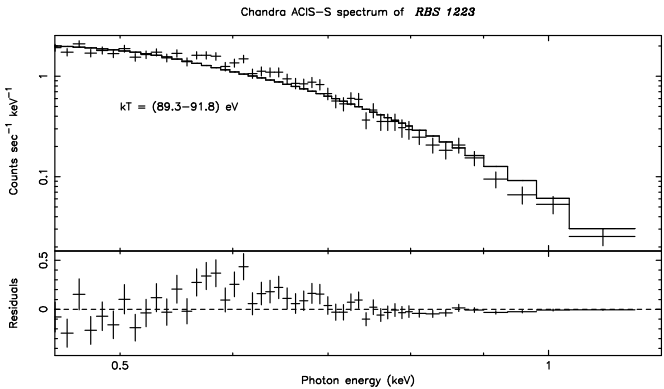}
\caption[ ]{A blackbody fit to the Chandra ACIS--S spectrum of RBS 1223.
A best--fit blackbody temperature is
 ${\rm kT}= (91 \pm 1) {\rm eV} $ 
($N_{\rm H} = (2.43 \pm 1.06) \times 10^{20}$~cm$^{-2} $).}
\label{chsp}
\end{figure}

With the fixed value of $N_{\rm H}$, the \chan\ ACIS--S spectrum of 
\ns\ can be represented by 
blackbody radiation with temperature $kT_{BB} = (90.6 \pm 0.8)$~eV 
with formal 1~$\sigma$ errors (Fig.~\ref{chsp}).

The total galactic column density in  the direction of RBS 1223 is 
$N_{\rm H, gal} = 2.1 \times 10^{20}$~cm$^{-2}$, 
 according to the Dickey \& Lockman, (\cite{nh}) maps of HI,  
averaged into $1^o \times 1^o$ bins
(routine $nh$ in FTOOLS).

\begin{figure}
\psfig{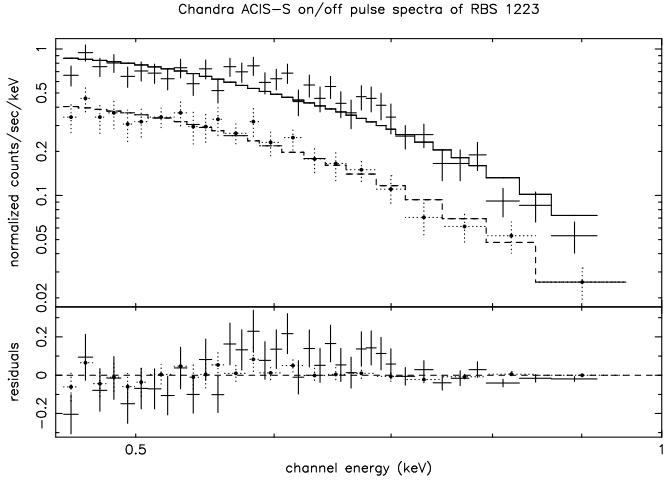}
\caption[ ]{The on/off pulse spectra of RBS 1223 obtained from Chandra ACIS-S
 observation. The on-pulse spectrum (upper) corresponds to the phases 0.2 to 
0.6 on Fig.~\ref{lces}, while the off-pulse spectrum (lower) is from 
averaging over the remaining phases. A best-fit blackbody temperatures are
 ${\rm kT}(on)= 93 \pm 3 {\rm eV}$ and ${\rm kT}(off)= 88 \pm 3 {\rm eV}$
 ($N_{\rm H} = 2.43 \pm 1.06 \times 10^{20}$~cm$^{-2} $).}
\label{phase_specs}
\end{figure}

In Fig.~\ref{phase_specs}, we show the ``on-pulse'' spectrum which is defined
by averaging over the phase  interval (0.2-0.6),  i.e. it includes 
the maximum of the light curve in Fig.~\ref{lces}. For comparison, we also 
show the ``off-pulse'' spectrum, obtained by averaging over the remaining 
phases. On- and off-pulse spectra were
rebinned such that each spectral bin contains at least 25 photons.
 A spectral fit, again with fixed column density 
$N_{\rm H} = 2.43 \times 10^{20}$~cm$^{-2} $, 
yield best-fit blackbody temperatures of
$kT {\rm (on)}= 93 \pm 3 {\rm eV}$ and $kT {\rm (off)}= 88 \pm 3 {\rm eV}$ 
with formal 3~$\sigma$ errors, indicating a trend
that the source is hotter at maximum phase (Fig.~\ref{phase_specs}).

\section{Discussion}

The detection of  a periodic signal with $P\simeq 5.16$~s
proves that \ns is a neutron star. Our determination of 
$\pdot \simeq (0.7 - 2.0) \times 10^{-11}~{\rm s s^{-1}}$
shows that \ns\ is spinning-down very rapidly.
Assuming a spin-down law of the type
$\fdot\sim - f^n$ ($f = 1/P$ is 
the pulsar rotation frequency, $\fdot$ is its derivative, and $n = \fddot
f/(\fdot)^2$ is the ``braking index'') with a
constant magnetic moment gives the age of the pulsating neutron star,
\be
\label{eq:tau}
\tau = \frac{P}{(n-1) \dot{P}} \left[ 1 - \left(\frac{P_0}{P},
\right)^{n-1} \right].
\ee
where $P_0$ is a period of the pulsar at birth.

Usually, an oblique rotating vacuum dipole
model is assumed, for which $n=3$ 
(Manchester \& Taylor \cite{mt77}). If $P_0$ is much smaller
than $P$, the expression for the age determination 
reduces to $\tau = P/(2 \pdot ) \equiv
\tau_{\rm c}$, the characteristic age of a pulsar. 

Thus, the characteristic age of \ns\ is 
$\tau_{\rm c} \simeq$ (6000 $-$ 12000)~yr.

Assuming  a neutron star radius of $10^6$\,cm and  
a moment of inertia of $10^{45}$\,g\,cm$^2$, 
the dipole magnetic field strength at the pole is 

\be
\label{eq:B}
B_{pole} = 6.4 \times 10^{19} \sqrt{P \dot{P}} \simeq (3.5 - 6.5) \times 10^{+14} \;\; {\rm G},
\ee

 well above  the so-called ``quantum critical field'' value 
$B_{\rm c} \equiv \frac{m_e^2 c^3}{e \hbar} = 4.4 \times 10^{13} \;\; {\rm G}$.
This indicates that \ns\ could be
 a {\it highly magnetized young} neutron star.

There has been growing recognition  in the literature of
a population of highly magnetized neutron stars. The circumstantial evidence
for this class comes from studies of soft gamma$-$ray repeaters (SGRs) and
anomalous \Xray\ pulsars (AXPs). 

SGRs are neutron stars showing multiple 
bursts of gamma--rays  with rather soft spectra
(e.g. Hurley~\cite{hurley2000} for a recent review).  The spin periods of SGRs are clustered in the 
interval 5--8 s. They all appear to be associated with supernova 
remnants, which limits their average age to approximately 
$20$ kyr. The angular offsets of the SGRs from 
the apparent centers of their associated supernova remnant shells 
indicate that SGRs are endowed with space velocities $>500$ km 
s$^{-1}$, greater than the space velocities of most 
radio pulsars (Cordes \& Chernoff~\cite{cordes98}). 

AXPs are similar to SGRs in that they are radio quiet X--ray pulsars 
with spin periods clustering in the range $6-12$ s, and have similar 
persistent X--ray luminosities as the SGRs ($\sim 10^{35}$ ergs 
s$^{-1}$, see e.g. Stella et al.~\cite{stella98}). Most of 
the AXPs appear to be associated with supernova remnants, and 
therefore they are also thought to be young neutron stars like 
the SGRs. The spin periods of both AXPs and SGRs are increasing with 
time (spinning--down), and show no evidence for intervals of decreasing 
spin period (spin--up), although the spin--down rates of many of the 
SGRs and AXPs appear to be variable or ``bumpy'' 
(e.g. Woods et al.~\cite{woods00}).

\begin{figure}
\centering
 \includegraphics[width=8cm]{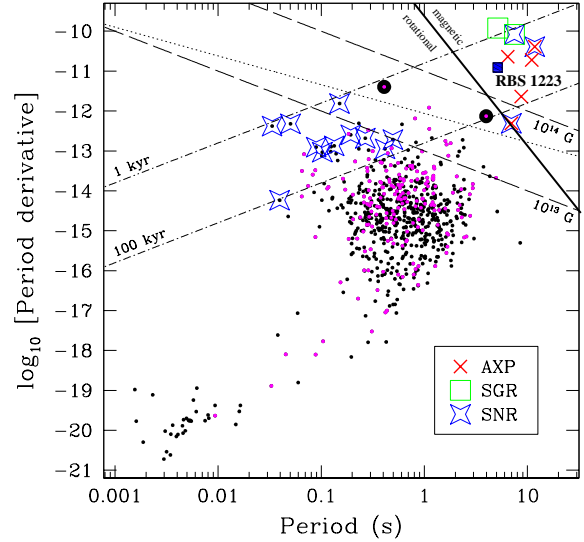}
      \caption{
Plot of $\pdot$ versus $P$ for radio pulsars (dots), anomalous X-ray
pulsars (AXPs), and soft gamma-ray repeaters (SGRs).  
The dotted line shown between the lines for $B=10^{13}$ and
$10^{14}$\,G indicates a hypothesized approximate theoretical boundary
(Baring \& Harding \cite{bh98b}) separating radio-loud and radio-quiet 
neutron stars due to effects relating to magnetic fields close to the 
critical field $B_{\rm c}$ (from Camilo et al. \cite{camilo}). 
\ns\ is identified by filled box. A solid line is indicating where
 magnetic field energy and rotational energy losses are equal 
($\edot_{mag}=-\frac{1}{6}\bdot_{pole}B_{pole}R^{3}$ and $\edot_{rot}=4\pi^2If\fdot$, Heyl \& Kulkarni \cite{heyl})}
   \label{ppdot}
   \end{figure}

In addition, it is worth noting that some of the SGRs and AXPs are 
also located well outside apparently associated supernova 
remnants (Marsden et al. \cite{marsdenetal}). Gaensler et al. 
(\cite{gaensleretal}) argue that they may not
be associated at all with these supernova remnants.   

 Our values of $P$ and $\pdot$ are suggestive, that \ns\ might belong to
either one of the two classes of unusual \Xray\ pulsars (Fig.~\ref{ppdot}).

 However, it is not excluded also that RBS 1223 may have
a conventional ($\sim 10^{12}$~G) magnetic field strength and identified 
as a neutron star in the propeller phase,
i.e. powered by inflow of matter from a disk around it (van Paradijs, Taam \& 
van den Heuvel \cite{vanPar}, Chatterjee, Hernquist, \& Narayan \cite{chatt},
Alpar \cite{alpar})

Due to the lack of an identification  at any other wavelength, 
the distance of RBS 1223  and hence its luminosity is quite uncertain. 

 Assuming a size of the \Xray\ emitting region of 3-10 km,
our best fit spectral parameters indicate a distance of 450-1500pc
($\sim $~1.2~kpc assumed by Motch (\cite{motch2000})). 

 Our determination of the absorption column density 
$N_{\rm H} = (0.6 - 6.6) \times 10^{20}~{\rm cm}^{-2} (3\sigma)$ 
(see Fig.~\ref{specs}) poses no further constraint on the distance of
RBS 1223.
The existence of diffuse high galactic latitude 
cirruses ( at the distance of 100$-$200 pc, 
see Magnani et al. \cite{magnani}) C9363 and C9383, which extend 
to the position of RBS 1223 (clearly seen in both IRAS 60$\mu$k and 
IRAS 100$\mu$k images, and having column  
density $N_{\rm H} \sim (3-4) \times 10^{20}$ cm$^{-3}$, 
Reach et al. \cite{reach})  suggests that \ns\ might be a 
nearby ($\sim$~100$-$200 pc)  or rather distant 
($\sim$~700$-$1500 pc) object.

However, the absence of any obvious supernova remnant in the general 
direction of this source (Green \cite{green}) and the fact that such a close
and young supernova remnant would not remain undetected in the X$-$rays, 
suggests the possibility that the source has a very large transverse 
velocity and it may have migrated far from its original position. 

In this  context it is worth  mentioning that RBS 1223 was not detected 
as a radio source during the VLA FIRST survey at 1.4 GHz 
(White et al.~\cite{first}) indicating a radio-quiet nature. 
The upper limit of radio flux at the source position is 0.94$\pm$0.138 mJy/beam.
  
 The position of \ns\  at the high 
galactic latitude  of  $b^{II}=+83^o$ imposes an additional 
restriction on its distance. 
If \ns\  would be at a distance $\sim 1000$~pc  and would be born
in the galactic plane, it must have traveled 
with a transverse velocity $v_t \sim 1000$~km~s$^{-1}$  
for about $t \sim 10^6$~yr. 
 On the other hand, blackbody temperatures in excess of 50~eV 
imply ages younger than $\sim 10^6$~yr for standard cooling curves 
(see Tsuruta \cite{tsuruta}). 
 This suggests a much smaller distance than 1~kpc. 

 Assuming a distance of the source of 
$\sim$~1000 pc and $kT_{BB} = 91 eV$, 
we determine the unabsorbed luminosity to be
\be
L_X \sim 4.1 \times 10^{32} {\rm ergs~s}^{-1}.
\ee

\section{Conclusions}

 We analyzed \chan\ ACIS--S  and an archival  \ros\ HRI 
observations of the isolated neutron star candidate \ns\ and 
found a period $P \sim 5.16$~s.  We also found that the neutron
star spins down at $\pdot \sim (0.7 - 2.0)\times 10^{-11}$
~s~${\rm s^{-1}}$  with some remaining uncertainty 
on the spin down rate. Confirmation of our result requires 
further X-ray observations of RBS 1223.

 The nature of RBS 1223 remains uncertain, even if the spin down of the pulsar can be confirmed independently.
It might be a young neutron star either with a very strong magnetic 
field (magnetar) or a neutron star with a more conventional magnetic 
field of $\sim 10^{12} {\rm G}$ in the propeller phase powered by 
inflow of matter from a surrounding disk. 

The absence of a nearby supernova remnant and the 
`characteristic age' of a pulsar suggest a high space velocity
to bring \ns\  away from its progenitor supernova remnant. 

Additional  constraints on its nature can be derived from detailed 
\Xray\ spectroscopy using upcoming XMM data and/or  an X-ray 
proper motion study which we have already initiated.

\begin{acknowledgements}
This project was supported by the 
Bundesministerium f\"ur Bildung und Forschung through 
the Deutsches Zentrum f\"ur Luft- und Raumfahrt e.V. (DLR) 
under grant number 50 OR 9706 8.
\end{acknowledgements}

{}

\end{document}